\documentclass[aps,prb,10pt,twocolumn,showpacs,preprintnumbers,amsmath,amssymb,]{revtex4-1} 

\usepackage{float}
\usepackage{graphicx}
\usepackage{tikz}
\usepackage{wasysym} % \hexagon symbol
\usepackage{dcolumn}
\usepackage{color}
\usepackage{amsmath}
\usepackage[breaklinks,colorlinks,bookmarks=false,citecolor=blue,linkcolor=red,urlcolor=blue]{hyperref}

\renewcommand{\vec}[1]{{\mathbf #1}}

\begin{document}

\title{Spin dynamics of the planar kagome lattice ferromagnet\\ with  four-site ring exchange processes}

\author{Jonas Becker}
\affiliation{Institut f\"ur Theoretische Festk\"orperphysik, JARA-FIT and JARA-HPC, RWTH Aachen University, 52056 Aachen, Germany}
\author{Stefan Wessel}
\affiliation{Institut f\"ur Theoretische Festk\"orperphysik, JARA-FIT and JARA-HPC, RWTH Aachen University, 52056 Aachen, Germany}
\date{\today}

\begin{abstract}
By means of quantum Monte Carlo simulations, combined with a stochastic analytic continuation, 
we examine the spin dynamics of the spin-1/2  planar (XY)  ferromagnet on the kagome lattice with  additional four-site ring exchange terms.  Such exchange processes were previously considered to lead into an extended $Z_2$ quantum spin liquid phase beyond a quantum critical point from the XY-ferromagnet. 
We examine the dynamical spin structure factor in  the non-magnetic regime and probe for signatures of spin fractionalization.
Furthermore, we contrast our findings and the corresponding energy scales of the excitation gaps in the ring exchange model to those emerging in a related Balents-Fisher-Girvin model with a  $Z_2$ quantum spin liquid phase, and monitor the softening of the magnon mode upon approaching the quantum critical point from the XY-ferromagnetic regime. 
\end{abstract}

\maketitle
Competing interactions  are an essential ingredient for stabilizing non-magnetic ground states of quantum magnets that exhibit topological order~\cite{Wen91a,Wen91b}. The exploration of such quantum spin liquid (QSL) states is thus at the forefront of current research in condensed matter physics~\cite{Balents10,Zhou17,Savary17}. 
For this purpose, it is particularly valuable that several model systems have been conceived, which were proven to stabilize gapped QSL states in extended parts of their ground state phase diagrams.  Prominent examples in this respect are Kitaev's toric-code and the honeycomb-lattice model~\cite{Kitaev97,Kitaev06}, as well as the kagome lattice-based  
Balents-Fisher-Girvin (BFG) model~\cite{Balents02}: For the latter system, the presence of an extended $Z_2$-QSL phase can be derived, e.g., from a perturbative consideration, in  which the bare spin-1/2 model maps onto an effective easy-axis spin-model with ring exchange processes on the kagome lattice's bow-ties, within a constrained low-energy manifold of states with zero total longitudinal magnetization on each (six-site) hexagon of the kagome lattice. 
The BFG model was furthermore shown to harbor fractionalized spin excitations and topological order in the QSL regime based on a formal equivalence to 
 a particular quantum dimer (QD) model~\cite{Balents02}. Such QD models have an exactly solvable limit, the Rokhsar-Kivelson point, at which the ground state is an exact superposition of all valid dimer coverings~\cite{Rokhsar88}, akin to a short-ranged resonating valence bond (RVB) state~\cite{Anderson73,Anderson87}. In the meantime, the characterization of the QSL phase in the  BFG model and variants thereof has been extended in various  aspects, including the topological   degeneracy and the corresponding contribution to the ground state entanglement, as well as the fractionalization of  magnetic excitations~\cite{Sheng05,Isakov06,Isakov07,Isakov11,Isakov12,Wang17,Sun18,Becker18}. 
Some of this progress was possible  because the BFG model can be examined by sign-problem free, unbiased quantum Monte Carlo (QMC)  simulations.

\begin{figure}[t]
	\includegraphics{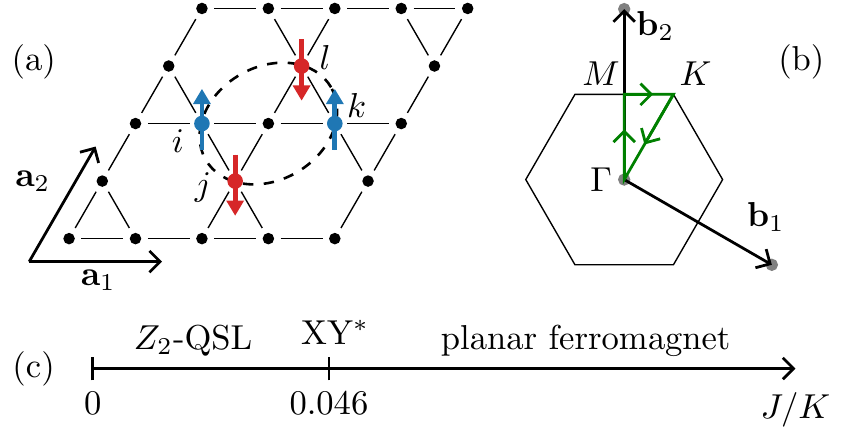} 
\caption{
(a) Kagome lattice with nearest neighbor (solid lines) and bow-tie ring exchange terms (dashed ellipse) of the $J$-$K$ model. 
The latter processes are indicated for a bow-tie with edges $i,j,k$ and $l$, along with a spin configuration that gets rotated by
the exchange term $P_{i,j,k,l}$. 
(b) First Brillouin zone (BZ) along with the path $\Gamma\rightarrow M\rightarrow K\rightarrow\Gamma$ (green lines).
(c) Ground state phase diagram of the $J$-$K$ model from  Ref.~\onlinecite{Dang11}.   
}\label{fig_sketch}	
\end{figure}

More recently, Dang, Inglis and Melko considered a spin-1/2 model with a direct competition between bare bow-tie ring exchange processes -- which emerged only  perturbatively  within the BFG  model -- and a ferromagnetic nearest neighbor transverse spin exchange~\cite{Dang11}.  More specifically, this $J$-$K$ model is described by the following spin-1/2 Hamiltonian, 
\begin{equation}
H  = -J\sum\limits_{\langle i,j \rangle} B_{i,j} - K \sum\limits_{\langle i,j,k,l \rangle} P_{i,j,k,l}.
\end{equation}
The summation of the transverse (XY) spin exchange, $B_{i,j} = S_i^+S_{j}^- + S_i^-S_{j}^+$, extends over the nearest neighbor bonds on the kagome lattice, and 
the  summation of the transverse ring exchange processes, $P_{i,j,k,l}= S_i^+S_j^-S_k^+S_l^-+S_i^-S_j^+S_k^-S_l^+$,   over all bow-ties on the kagome lattice, cf. Fig.~\ref{fig_sketch}(a) for an illustration. This model  can be studied by sign-problem-free QMC simulations  for positive values of $J$ and $K$, using a multi-branch cluster update scheme~\cite{Sandvik99,Melko05}. 
For $K=0$, the ground state is a planar XY-ferromagnet, i.e., a superfluid in the equivalent hard-core boson representation.  The ring exchange  processes reduce  the ferromagnetic alignment, and  the authors of Ref.~\onlinecite{Dang11} identified a continuous quantum phase transition at a critical value of $K/J\approx 21.8$ ($J/K\approx 0.046$) to a large-$K$ phase,  which exhibits no apparent symmetry breaking, cf. Fig.~\ref{fig_sketch}(c). By analogy to the BFG model, they concluded that (i) the large-$K$ regime of the $J$-$K$ model is a gapped $Z_2$-QSL phase, and (ii) the quantum phase transition is a candidate for  the  XY${}^{*}$ universality class, in view of the  fractionalization of the magnetic excitations within  $Z_2$-QSLs and the condensation
of the spinons at the transition  (a correspondingly  large value of the anomalous exponent $\eta$ was  not determined in Ref.~\onlinecite{Dang11}). 
Regarding the correspondence between the $J$-$K$  and the BFG model, we note that (i) the ring exchange  is an emerging low-energy process in the BFG model, while it is the leading term in the large-$K$ regime of the $J$-$K$ model,
(ii)  the $J$-$K$ model does not feature an explicit constraint on the hexagonal magnetizations -- within the QD model description, this allows for  monomer defects  to enter the low-energy manifold of the $J$-$K$ model in addition to  perfect dimer coverings. 
In view of these differences, it  is thus worthwhile to examine the $J$-$K$ model using  probes that allow us
to gain additional insight into the  nature of its large-$K$ regime. A powerful diagnostic approach to identify possible QSL phases are spectroscopic measurements probing, e.g.,  for scattering continua from fractionalized spin excitations. 

Here, we  thus examine in particular the  spin dynamics of the $J$-$K$ model in terms of the dynamical spin structure factor (DSSF), in order to probe for signatures of QSL physics in the large-$K$ regime. 
We profit from recent QMC studies of the DSSF  of the BFG model, which explicitly uncovered the fractionalized magnetic excitations in the established QSL regime~\cite{Sun18,Becker18}. 
Below, we  therefore address  also  the similarities and differences between these two models.  In addition, we  study the evolution of the DSSF upon varying the ratio $J/K$.

\begin{figure}[t]
\includegraphics[width=0.5\textwidth]{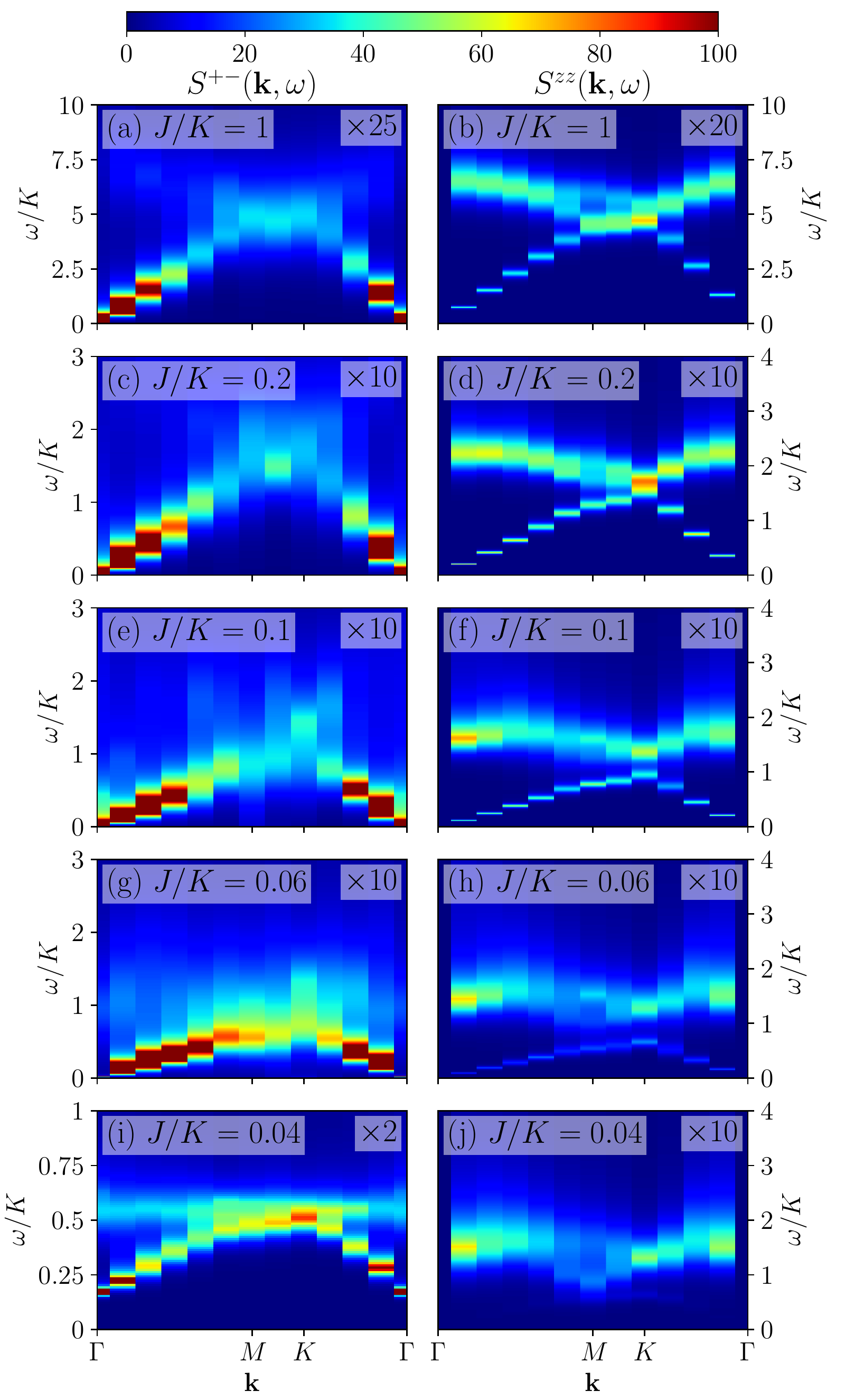}
\caption{
DSSF $S^{+-}(\vec{k},\omega)$ (left panels) and $S^{zz}(\vec{k},\omega)$ (right panels) of the $J$-$K$ model along the BZ path
$\Gamma\rightarrow M\rightarrow K\rightarrow\Gamma$  [cf. Fig.~\ref{fig_sketch}(b)] for different ratios $J/K$ from QMC simulations. 
To fit to a common scale, the intensities were multiplied by individual factors, which are provided in the upper right corner of each panel separately.
}\label{fig_all}
\end{figure}

To calculate the DSSF of the $J$-$K$ model, we performed QMC simulations using the sampling method
from Ref.~\onlinecite{Becker18}:
We consider finite rhombic systems with $N_s=3L^2$ lattice sites and periodic boundary conditions along both  lattice directions $\mathbf{a}_1$, $\mathbf{a}_2$ in Fig.~\ref{fig_sketch}(a), 
with the unit cell distance fixed to  $a=1$. 
To formulate the DSSF on the three-sublattice kagome lattice, we
denote by $\vec{S}_{i,\alpha}$  the spin at position $\vec{r}_{i,\alpha}$ on sublattice $\alpha$ ($=1,2,3$) in the $i$-th unit cell ($i=1,...,L^2$).
We then obtain $3\times 3$ correlation-matrices 
$S^{+-}_{\alpha,\beta}(\vec{k},\omega) = \int\mathrm{d}t \, \mathrm{e}^{-\mathrm{i}\omega t} \langle {S}^+_{\vec{k},\alpha}(t) {S}^-_{-\vec{k},\beta}(0) +  {S}^-_{\vec{k},\alpha}(t) {S}^+_{-\vec{k},\beta}(0)\rangle$
for the transverse,  and
$S^{zz}_{\alpha,\beta}(\vec{k},\omega) = \int\mathrm{d}t \, \mathrm{e}^{-\mathrm{i}\omega t} \langle {S}^z_{\vec{k},\alpha}(t) {S}^z_{-\vec{k},\beta}(0)  \rangle$
for the longitudinal 
channel respectively, where
$\vec{S}_{\vec{k},\alpha}=(1/L) \sum_i \mathrm{e}^{-\mathrm{i} \vec{k}\cdot \vec{r_{i,\alpha} }} \vec{S}_{i,\alpha}$. We examine separately the traces over the  correlation-matrices in each channel, 
$S^{+-}(\vec{k},\omega) :=\sum_\alpha S^{+-}_{\alpha,\alpha}(\vec{k},\omega)$, and 
$S^{zz}(\vec{k},\omega) :=\sum_\alpha S^{zz}_{\alpha,\alpha}(\vec{k},\omega)$, respectively, which contain the sum over the correlation-matrix eigenvalues of the spectral functions at each fixed momentum transfer   $\vec{k}$ from the first Brillouin zone (BZ).
The BZ is shown in terms of the  reciprocal lattice vectors $\vec{b}_1$,  $\vec{b}_2$  in Fig.~\ref{fig_sketch}(b). 
In the following, we show QMC data for systems with $L=12$. Simulations performed for $L=18$ at selected parameter sets returned similar results. To target the ground state regime, the temperature $T$ was adapted to ${J}/({2L})$ accordingly~\cite{Sun18,Becker18}, and  simulated annealing was used during the initial  stage of the thermalization phase to improve equilibration.
In order to  extract the DSSF
from the QMC simulations, we  measured the  transverse imaginary-time displaced spin-spin correlation functions and accessed the longitudinal correlations directly in Matsubara frequency space~\cite{Michel07,Michel07a},  using the stochastic analytic continuation method in the formulation of Ref.~\onlinecite{Beach04} to obtain the  
spectral functions in real frequencies for both cases. 
We  performed the analytic continuations independently for the three correlation-matrix eigenvalues for each given momentum  $\vec{k}$, in order to enhance the spectral resolution as detailed in Ref.~\onlinecite{Becker18}.

We begin the investigation of the DSSF in the XY-ferromagnetic regime of the $J$-$K$ model, cf. Fig.~\ref{fig_sketch}(c). The evolution of the DSSF upon varying  $J/K$ is shown in 
Fig.~\ref{fig_all}. For $J/K=1$, (top panels) we  observe the characteristic features of the XY-ferromagnetic regime (similar to the $J$-only limit ($K=0$), cf. Ref.~\onlinecite{Becker18}): In the transverse channel, $S^{+-}(\vec{k},\omega)$, the  spectral weight is dominantly contributed by a gapless low-energy magnon mode, i.e., the Goldstone soft-mode from the $U(1)$ symmetry breaking. 

While the Goldstone mode  contributes also some (weaker) spectral weight to $S^{zz}(\vec{k},\omega)$, the longitudinal channel is dominated by  optical magnons  above $\omega\approx 5K$, with a dispersion maximum at the 
$\Gamma$-point. Upon reducing $J/K$, the overall energy scale of these excitations  decreases correspondingly, cf. Fig.~\ref{fig_all},  along with a reduction of the spin wave velocity of the lower magnon mode (estimated from the slope of the Goldstone mode near the $\Gamma$-point). The bottom panels of Fig.~\ref{fig_all}, for $J/K=0.04$, reside beyond the quantum phase transition, and exhibit finite excitation gaps in both channels, in accord with a gapped, non-ferromagnetic ground state. 

\begin{figure}[t]
\includegraphics[width=0.5\textwidth]{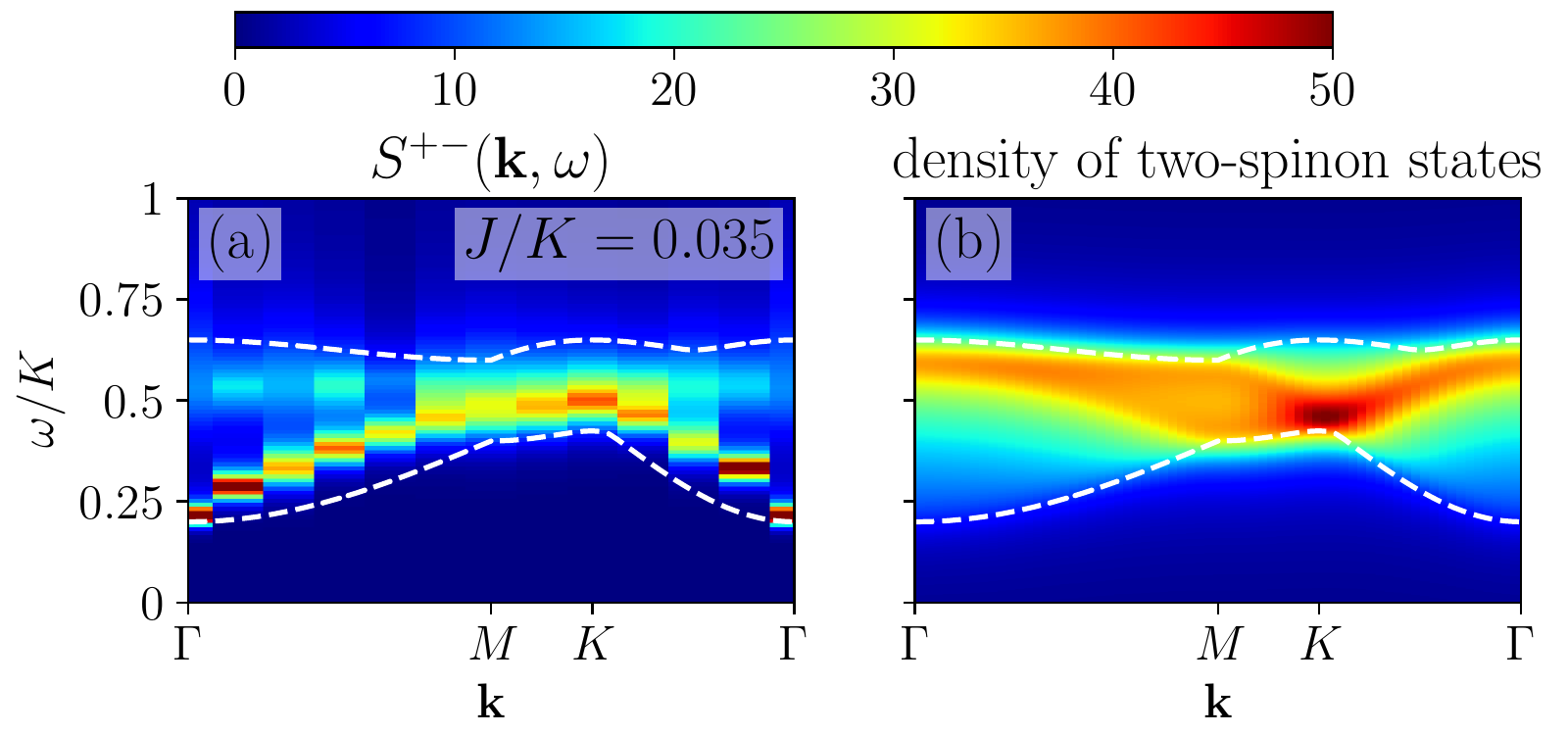}
\caption{
(a) DSSF $S^{+-}(\vec{k},\omega)$ of the $J$-$K$ model for   $J/K=0.035$ 
along the BZ path $\Gamma\rightarrow M\rightarrow K\rightarrow\Gamma$. 
(b) Density of two-spinon states within the tight-binding model with $t=0.025K$, and $\Delta_s=0.25K$.
In both panels,  dashed lines  indicate the lower and upper threshold of the two-spinon continuum within the tight-binding model.  
}\label{fig_QSL}
\end{figure}

The transverse channel for $J/K=0.035$, which resides even further beyond the quantum critical point, is shown in Fig.~\ref{fig_QSL} (a).  Apart from a slightly larger gap of about $\Delta_\mathrm{T}\approx 0.24 K$, it features a very similar overall structure as the DSSF for $J/K=0.04$: The spectral weight is dominantly concentrated along the lower edge of the spectral support. This  differs from the more broadly extended distribution of the spectral weight in the DSSF for the QSL regime of the BFG model~\cite{Sun18, Becker18}: in the latter case, the transverse DSSF indeed compares well to a tight-binding model of two-spinon continuum states~\cite{Becker18}, providing a smoking-gun signature for fractionalized spin excitations.

For a direct comparison to the $J$-$K$ model data, the two-spinon DSSF is shown in Fig.~\ref{fig_QSL}(b).
For the latter, the  two-spinon contribution to $S^{+-}(\vec{k},\omega)$  
is defined in terms of a tight-binding model that treats spinons as free particles~\cite{Becker18}, with a dispersion relation $\epsilon_{\vec{k}}=\Delta_s+\epsilon_t(\vec{k})$, where $\Delta_s$ 
quantifies the local energy cost of a single spinon, and
$\epsilon_t(\vec{k})=-2t[\cos(\vec{a_1}\vec{k})+\cos(\vec{a_2}\vec{k})+\cos(\vec{a_2}\vec{k}-\vec{a_1}\vec{k})]$
the triangular lattice tight-binding dispersion with $t\sim J$ the nearest-neighbor hopping, fitted to the bandwidth of the QMC spectrum. 
In this approximation, $S^{+-}(\vec{k},\omega)\approx \frac{4\pi}{L^2} \sum_{\vec{k}'} \delta(\omega-\epsilon_{\vec{k}'}-\epsilon_{\vec{k}-\vec{k}' })$
equals the  density of two-spinon states, for which interaction effects have been  accounted for phenomenologically by a Lorentzian $\delta$-function broadening~\cite{Kourtis16,Huang17,Becker18}. Whereas the spectral support of  $S^{+-}(\vec{k},\omega)$ in Fig.~\ref{fig_QSL}(a) can be  captured by the two-spinon model, the actual spectral weight distributions differ substantially: In particular, as already mentioned above, in the $J$-$K$ model the dominant spectral weight resides along the bottom edge of the spectral support, whereas the two-spinon density of states  gets suppressed in this regime and instead concentrates more towards the upper edge. While $S^{+-}(\vec{k},\omega)$ in Fig.~\ref{fig_QSL}(a) also exhibits some spectral weight in this regime, we conclude that the spinon tight-binding model, which captures the transverse DSSF of the BFG model within the $Z_2$-QSL phase rather well, is not  particularly  adequate for the $J$-$K$ model within the anticipated QSL regime. 

\begin{figure}[t]
\includegraphics[width=0.5\textwidth]{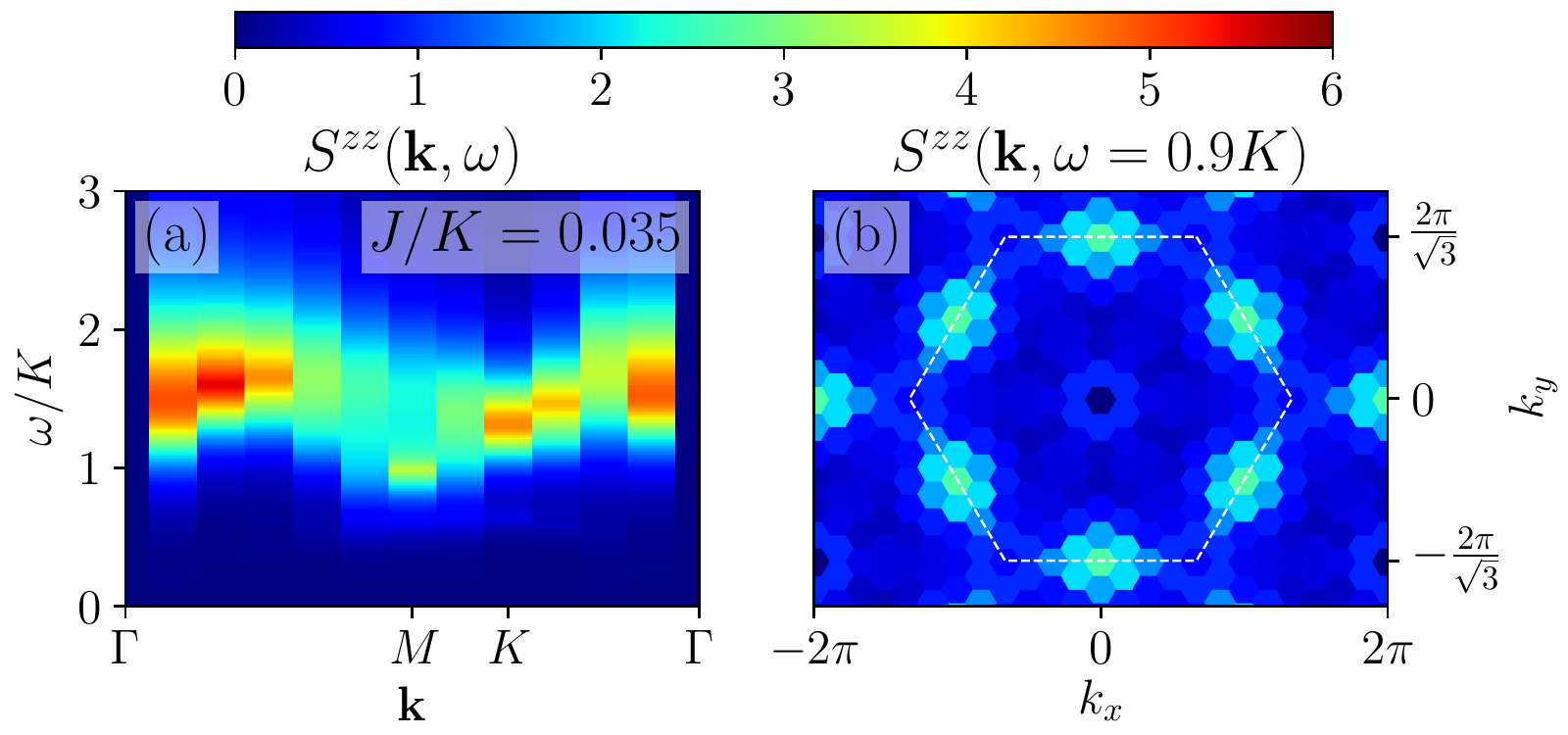}
\caption{
DSSF $S^{zz}(\vec{k},\omega)$ of the $J$-$K$ model for $J/K=0.035$,
(a) along the BZ path $\Gamma\rightarrow M\rightarrow K\rightarrow\Gamma$, 
and (b) at $\omega=0.9K$. The  hexagon in (b) denotes the BZ.  
}\label{fig_Szz}
\end{figure}

Upon examining the longitudinal channel, we observe   more similarities   in  $S^{zz}(\vec{k},\omega)$ between the $J$-$K$ model and the BFG model QSL, than in the transverse channel: 
In Fig.~\ref{fig_Szz}(a), for $J/K=0.035$, the excitation gap in the longitudinal channel, $\Delta_\mathrm{L}\approx K$, is
controlled by the ring exchange energy scale, which is an explicit term in the bare $J$-$K$ model Hamiltonian. In fact, for the $Z_2$-QSL phase of the BFG model, the gap in $S^{zz}(\vec{k},\omega)$ is also of the order of the strength of the (emerging) ring exchange processes~\cite{Sun18, Becker18}. However, due to the explicit presence of the ring exchange term in the $J$-$K$ model, the longitudinal gap $\Delta_\mathrm{L}$ is in this case larger than $\Delta_\mathrm{T}$, whereas this relation is inverted for the BFG model, for which instead $\Delta_\mathrm{L}$ defines the lowest excitation gap. 

The spectral weight distribution in Fig.~\ref{fig_Szz}(a) is overall remarkably  similar to the BFG model  result (cf. Ref.~\onlinecite{Becker18}): In particular, we identify a dip in  $S^{zz}(\vec{k},\omega)$ at the $M$ point of the BZ, which is seen more explicitly in the constant energy cut in Fig.~\ref{fig_Szz}(b) at $\omega=0.9K$. 
For the BFG model, this repeating structure at the $M$ points was anticipated in Refs.~\onlinecite{Sun18,Becker18} to provide a fingerprint for crystal momentum fractionalization~\cite{Wen02,Cheng16,Essin14} of the vison  excitations within the $Z_2$-QSL state, with the corresponding value of $\Delta_\mathrm{L}$ providing the two-vison excitation gap. A similar interpretation in the present case would locate the vison excitations out of the low-energy regime (as $\Delta_\mathrm{L}>\Delta_\mathrm{T}$). 

\begin{figure}[t]
\includegraphics[width=0.5\textwidth]{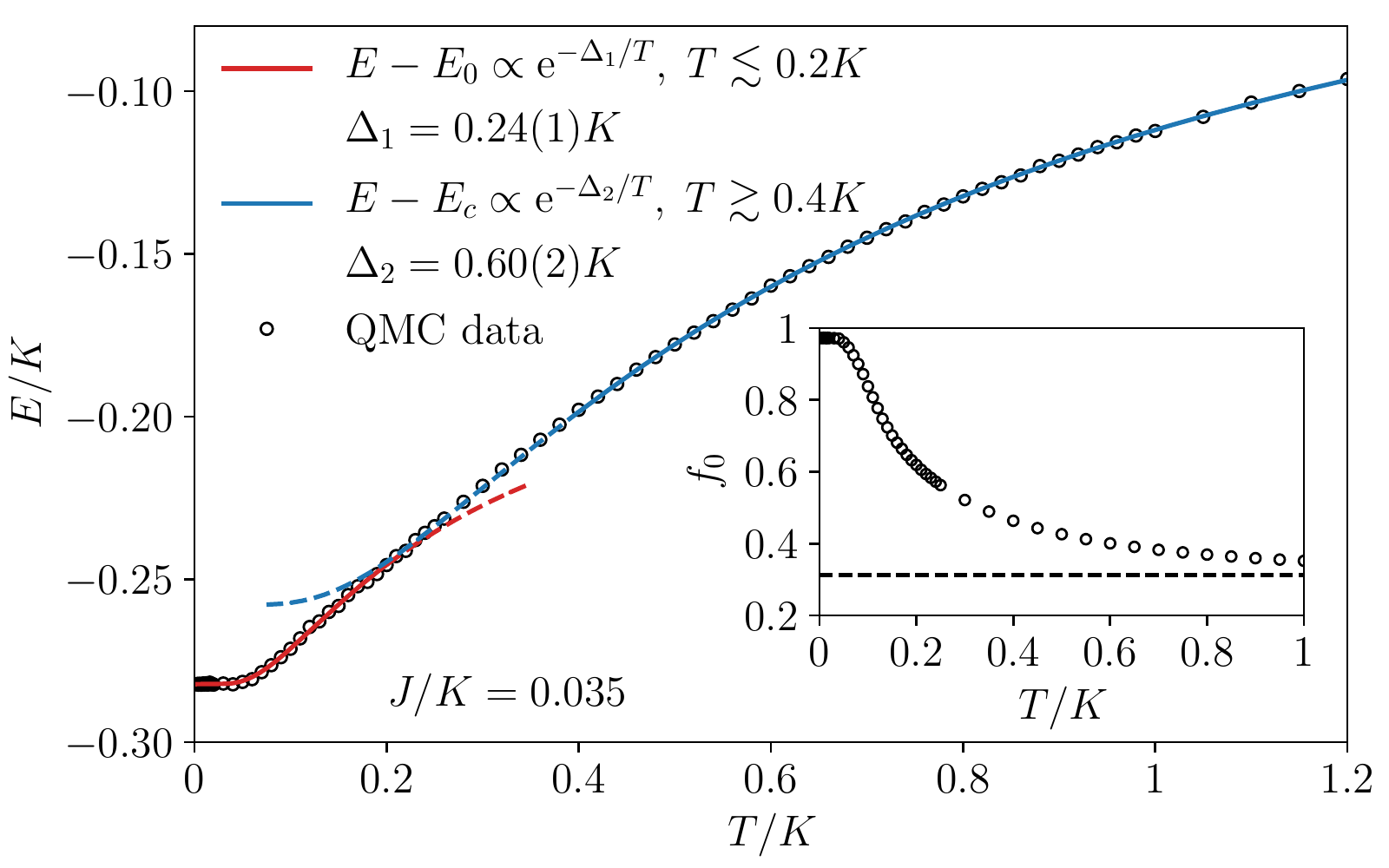}
\caption{
Temperature dependence of the internal energy $E$ of the $J$-$K$ model for $J/K=0.035$, along with exponential fits to activated behavior atop the ground state energy $E_0$
%$E_0\approx -0.2822K$ 
(red line) and the crossover energy   
 $E_c\approx -0.258K$ (blue line). The inset shows the temperature dependence of the mean fraction $f_0$ of hexagons that satisfy the QD model subspace constraint on the longitudinal magnetization. The dashed line indicates the asymptotic  value of $0.3125$.
}\label{fig_E}
\end{figure}

This hierarchy of energy scales is  also imprinted in the temperature dependence of the internal energy $E$, shown in Fig.~\ref{fig_E} for $J/K=0.035$. In contrast to the BFG model~\cite{Isakov07, Sun18,Becker18}, we do not observe an extended paramagnetic plateau in the temperature dependence of $E$ 
in Fig.~\ref{fig_E}, cf.  the less pronounced difference in the two excitation gaps $\Delta_\mathrm{L}$, and $\Delta_\mathrm{T}$ in this case. Nevertheless, we can discern the observed temperature dependence in terms of two sectors:  A low-$T$ activated behavior atop the ground state energy due to the thermal proliferation of excitations with a gap of order $\Delta_\mathrm{T}$, and a higher-$T$ contribution beyond a crossover  energy $E_c$, with a larger gap of the order of $\Delta_\mathrm{L}$, cf. the fits in  Fig.~\ref{fig_E}. 

In the inset of Fig.~\ref{fig_E}, we finally examine the temperature dependence of the mean fraction $f_0$ of hexagons on the kagome lattice, for which the constraint $S^z_{\hexagon}=0$ on the longitudinal magnetization ($\hexagon$ denotes one of the six-site hexagons, and $S^z_{\hexagon}=\sum_{j\in\hexagon} S^z_j$) --  defining the QD model subspace in the BFG model -- is satisfied. For large temperatures, $f_0$ approaches the unconstrained value of ${6 \choose 3}/2^6=0.3125$. The  fraction  $f_0$ increases steadily upon lowering $T$, and in the low-temperature regime,  well below the gap scales,  it saturates to about 97\%. The residual fluctuations  from the QD model subspace are  due to   spin exchange processes  mediated by the two-site ($J$) exchange terms.  The $J$-$K$ model thus  dynamically generates the QD model subspace constraint effectively, even though an explicit penalty term, such as  $\sum_{\hexagon} (S^z_{\hexagon})^2$, is absent in its Hamiltonian (here, the summation is performed over all the hexagons of the kagome lattice).

In conclusion, we examined the DSSF of the $J$-$K$ model in the two distinct regions of the low-$K$ XY-ferromagnetic, and the large-$K$ non-magnetic regime. For the latter, we identified two distinct excitation gaps, of which the larger one relates  to the energy scale of the bare ring exchange processes of the $J$-$K$ model.  Moreover, the overall structure in the longitudinal channel resembles the DSSF in the QSL regime of the BFG model, hence providing spectral support for the identification in Ref.~\onlinecite{Dang11} of a corresponding QSL state in the $J$-$K$ model. However, our results for the transverse DSSF do not  further strengthen  this scenario, as the spectral weight distribution for the $J$-$K$ model  shows  qualitatively distinct features from the two-spinon continuum that is characteristic for the QSL state of the BFG model. In order to perpetuate the case for a QSL phase in the $J$-$K$ model, it will thus be important to provide a compelling description of the transverse DSSF in terms of fractionalized spin excitations that accounts on a quantitative level for the QMC data that we reported here. We  hope that our findings will motivate further investigtions in this direction.

\textit{Acknowledgements}.
We thank  M. Loh\"ofer, R. G. Melko, and F. Pollmann for useful discussions.
Furthermore, we acknowledge support by the Deutsche Forschungsgemeinschaft (DFG) 
under grant FOR 1807 and RTG 1995, and thank the IT Center at RWTH Aachen University 
and the JSC J\"ulich for access to computing time through JARA-HPC.
%

%
% ----bib has been sorted according to citation order----
%
\end{document}